\documentclass[aip,jap,reprint,twoside,floatfix,showpacs,showkeys]{revtex4-1}

\usepackage[utf8]{inputenc}
\usepackage[T1,T2A]{fontenc}
\usepackage{amsmath,amssymb,amsfonts}
\usepackage{latexsym,eucal,gensymb}
\usepackage{dcolumn}
\usepackage{graphicx}
\usepackage{subfig}
\usepackage{bm}
\usepackage[version=3]{mhchem}
\usepackage{miller}
\usepackage{textcomp}
\usepackage{array}
\usepackage{xspace}
\usepackage{color}
\usepackage{exscale}
\usepackage{xcolor}
\usepackage{dcolumn}
\usepackage{physics}

\usepackage{ifpdf}
\ifpdf
\usepackage[pdftex,unicode,colorlinks,allcolors=blue,breaklinks,baseurl=http://]{hyperref}
\hypersetup{
	pdfauthor   = {Damir Islamov},%
	pdftitle    = {The charge transport mechanism and electron trap nature in thermal oxide on silicon},%
	pdfsubject  = {SILC},%
	pdfkeywords = {silica, oxygen vacancy, traps, charge transport, leakage current},%
	pdfcreator  = \LaTeX,%
	pdfproducer = \LaTeX}
\else
\usepackage[dvips,unicode,colorlinks,allcolors=blue,breaklinks,baseurl=http://]{hyperref}
\fi

\newcommand{\Wt}{\ensuremath{W_\mathrm{t}}}
\newcommand{\Wopt}{\ensuremath{W_\mathrm{opt}}}

\newcommand{\m}{\ensuremath{m^*}}
\newcommand{\me}{\ensuremath{m_\mathrm{e}}}
\newcommand{\nt}{\ensuremath{n_\mathrm{t}}}

\begin{document}
\title{The charge transport mechanism and electron trap nature in thermal oxide on silicon}

\author{Damir~R. Islamov}\email{damir@isp.nsc.ru}
\affiliation{Rzhanov Institute of Semiconductor Physics,
	Siberian Branch of the Russian Academy of Sciences,
	Novosibirsk 630090, Russian Federation}%
\affiliation{Novosibirsk State University,
	Novosibirsk 630090, Russian Federation}%
\author{Vladimir~A. Gritsenko}
	\affiliation{Rzhanov Institute of Semiconductor Physics,
		Siberian Branch of the Russian Academy of Sciences,
		Novosibirsk 630090, Russian Federation}%
\affiliation{Novosibirsk State University,
		Novosibirsk 630090, Russian Federation}%
\author{Timofey~V. Perevalov}
\affiliation{Rzhanov Institute of Semiconductor Physics,
	Siberian Branch of the Russian Academy of Sciences,
	Novosibirsk 630090, Russian Federation}%
\affiliation{Novosibirsk State University,
	Novosibirsk 630090, Russian Federation}%
\author{Oleg~M. Orlov}\email{oorlov@mikron.ru}
		\affiliation{JSC Molecular Electronics Research Institute, Zelenograd, Moscow 124460, Russian Federation}%
\author{Gennady~Ya. Krasnikov}
	\affiliation{JSC Molecular Electronics Research Institute, Zelenograd, Moscow 124460, Russian Federation}%

\date{\today}

\begin{abstract}
	The charge transport mechanism of electron via traps
	in amorphous \ce{SiO2} has been studied.
	Electron transport is limited by phonon-assisted
	tunneling between traps.
	Thermal and optical trap energies
	$\Wt=1.6$\,eV, $\Wopt=3.2$\,eV, respectively, were determined.
	Charge flowing leads to oxygen vacancies generation,
	and the leakage current increases due to
	the increase of charge trap density.
	Long-time annealing at high temperatures
	decreased the leakage current to initial values
	due to oxygen vacancies recombination with
	interstitial oxygen.
	It is found that the oxygen vacancies act as electron traps in \ce{SiO2}.
\end{abstract}


\maketitle


Silica \ce{SiO2} is the key material in electronic and optical devices, fibers. 
Intrinsic defects in \ce{SiO2} act as localization centers
for electrons and holes (traps).
The presence of such defects in particular layers of a device
results in the whole device degradation.
The main intrinsic defects in \ce{SiO2} are
threefold coordinated silicon atom with
an unpaired electron (\ce{#Si.}) \cite{a-SiO2:E1Center:PLR78:887,SiO2:S-O:PLR114:115503,H-VO-SiO2:PhysRevB.92.014107},
the oxygen vacancy (\ce{Si-Si} bond)
\cite{Defects:SiO2:PhysRevB.27.3780,SiO2:VO:prl:97:066101,SiO2:opt+epr-defects:prb75:024109,SiO2:VO:prb67:033202,SiO2:luminescence2.7eV:PhysRevLett.62.1388,a-SiO2:VO:luminescence:PhysRevLett.79.753},
non-bridging oxygen (\ce{#Si-O.})
\cite{Defects:SiO2:PhysRevB.27.3780, SiO2:e-trap:prl99:136801},
peroxide radical (\ce{#Si-O-O.})
\cite{Defects:SiO2:PhysRevB.27.3780, SiO2:opt+epr-defects:prb75:024109}
and
a peroxide bridge (\ce{#Si-O-O-Si#})
\cite{Defects:SiO2:PhysRevB.27.3780}.
Here the sign (\ce{-}) shows a chemical bond
formed by two electrons,
the sign (\ce{.}) denotes an unpaired electron.
On the base of numerous experiments it was found
that oxygen vacancies in \ce{SiO2} act as traps for holes
\cite{SiO2:opt+epr-defects:prb75:024109,a-SiO2:VO:luminescence:PhysRevLett.79.753,SiO2+SixNy:traps:CRSSMS36:129}.
According to the results of quantum-chemical simulations,
it was  predicted that the oxygen vacancy
can act as traps for electrons \cite{SiO2:e-trap:prl99:136801, SiO2:VO:mee80:292}.
However, the electron trapping energy
and the trap ionization mechanism
in strong electric field are unknown.

Long-time charge flowing through \ce{SiO2} in a strong electric field
($F\approx 10^7$\,V/cm) leads to the oxide conductivity
increase at lower electric field values
($\approx 2\times 10^6$\,V/cm).
Thus, an extra current through the dielectric is added to
the Fowler-Nordheim tunneling current
\cite{Fowler-Nordheim, Sze:Devices:Wiley:2006},
and addition values are comparable to the tunneling ones.
This phenomenon is called Stress Induced Leakage Current (SILC)
\cite{Defects:SiO2:PhysRevB.27.3780,SiO2:VO:prl:97:066101,SiO2:opt+epr-defects:prb75:024109,SiO2:VO:prb67:033202}.
Despite the fact that SILC is studied widely,
both experimentally and theoretically
\cite{SiO2:SILC:PRL83:372:1999, SiO2:SILC:JAP86:2095:1999, SiO2:TAIT-SILC:JAP90:3396:2001, SiO2:SILC:JAP92:2593:2002},
the nature of the defects responsible for SILC and
charge transport mechanism are still debatable questions.
The purposes of the present study are
identification of the ionization mechanism of electron traps
in \ce{SiO2} in external electric field
and determination of the trap parameters
(ionization energy, concentration).
The phenomenon of SILC was used to generate the traps in \ce{SiO2}
followed by a charge transport study.

SILC and transport measurements were performed on
FET transistors with the floating gate,
manufactured using the $180$\,nm design rule technology.
The $p$-\ce{Si} substrate was used as the bottom contact,
the $n^+$-type \textsl{poly}-\ce{Si} floating gate was used
as the top contact.
The tunnel \ce{SiO2} layer thickness was $7.5$\,nm.
Test samples were formed of $16$ parallel-connected
$n$-\ce{Si}/\ce{SiO2}/\textsl{poly}-\ce{Si} capacitors
with the total area of the \textsl{poly}-silicon electrodes
of $8\times 10^4$\,$\mu$m$^2$.
SILC was generated by the current of $1$\,mA/cm$^2$.
The total value of the flown charge via tunnel \ce{SiO2}
was $0.01$--$10$\,C/cm$^2$.
Transport measurements have been performed at temperatures of $25$--$70$\,\celsius\xspace.

Simulation was conducted for a 72-atom $\alpha$-\ce{SiO2} supercell
with the Quantum-ESPRESSO software within the scope of
the density functional theory
with B3LYP hybrid functional \cite{QE-2009}.
This method yields the $\alpha$-\ce{SiO2} band gap of $8.0$\,eV,
in excellent agreement with the experimental value
\cite{SiO2:PE-electrons:pr140:1965}
and with the theoretical value of $8.1$\,eV
obtained for amorphous \ce{SiO2} within framework
of similar approach \cite{H-VO-SiO2:PhysRevB.92.014107}.
See supplemental material at \cite{silc-supp-mat} for
detailed description of the numeric simulation procedure.
 [give brief description of material].
Electron and hole trap energies $\Wt^\mathrm{e}$ and
$\Wt^\mathrm{h}$ at the \ce{Si-Si} bond are evaluated
as difference between perfect and defect supercell
electron affinities and ionization energies as follows:
\begin{equation}
	\Wt^\mathrm{e/h}=E_\mathrm{p}^{+1/-1}-E_\mathrm{d}^{+1/-1}-E_\mathrm{p}^{0}+E_\mathrm{d}^{0}.
	\label{e:Wt:eh}
\end{equation}
Here $E_\mathrm{p}^{0/+1/-1}$ and $E_\mathrm{d}^{0/+1/-1}$
represent the values for the total energy of an ideal and
defect supercell with a different amount of the total charge.
This approach was proposed for \ce{SiO2}
to obtain the defect level positions between different charge states
\cite{SiO2:defects:prb41:5061},
and used for charge localization energy calculation
in \ce{Si3N4} \cite{Si3N4:DFT-traps:apl89:2006}
and \ce{HfO2} \cite{Foster:PRB65:174117}.

\begin{figure}[tb]
	\includegraphics[width=\columnwidth]{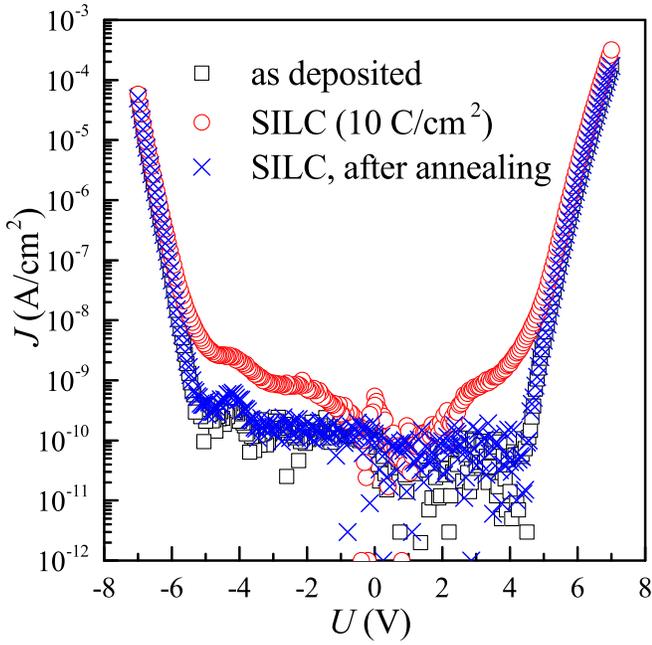}
	\caption{(Color online)
		Experimental current-voltage characteristics
		of $n$-\ce{Si}/\ce{SiO2}/\textsl{poly}-\ce{Si} structures
		at room temperature ($25$\,\celsius\xspace)
		before SILC ($\Box$), after SILC ($\circ$), and
		after SILC followed by long-time annealing ($\times$).}
	\label{f:IV25:annealed}
\end{figure}

Experimental current-voltage characteristics ($J$-$V$)
of the as deposited oxide are shown in Fig.~\ref{f:IV25:annealed}
by box symbols.
At high voltages $\abs{U}>5$\,V ($F>6$\,MV/cm) the current depends on
the voltage exponentially and is limited by
the tunnelling emission that proceeds
by the Fowler–Nordheim mechanism \cite{Fowler-Nordheim}:
\begin{equation}
	\begin{array}{c}
		J_\mathrm{FN} = A F^2  \exp\left(-\frac{B\sqrt{\m}\Phi_0^{3/2}}{F} \right), \\
		A = \frac{e^3}{8\pi h \Phi_0}, B = \frac{8\pi\sqrt{2}}{3he},
	\end{array}
	\label{e:J:Fowler-Nordheim}
\end{equation}
where $J_\mathrm{FN}$ is the tunneling current density,
$\m$ is the electron effective tunneling mass,
$\Phi_0$ is the height of the triangular potential barrier
for electrons at the \ce{Si}/\ce{SiO2} interface,
$e$ is the elementary (electron) charge,
$h$ is the Planck constant.
When $\abs{U}<5$\,V, the measured current values will be
determined by the sensitivity of the measuring devices and
bulk properties of the substrate.
According to photo emission measurements,
the height of the triangular potential barrier
for electrons at the \ce{Si}/\ce{SiO2} interface is
$\Phi = 3.14$\,eV \cite{Si:SiO2:photoemission:1978:en}.
The electron energy spectrum quantization in high fields
at the \ce{Si}/\ce{SiO2} interface leads to a descrease of
the effective barrier height to $\Phi_0 = 2.9$\,eV \cite{SiO2:FN:Barriers:JAP53:5052:1982}.
Taking $\Phi_0 = 2.9$\,eV for simulations of experimental data
with Eq.~(\ref{e:J:Fowler-Nordheim}), we get
the electron effective mass
in \ce{SiO2} $\m/\me=0.5\pm 0.02$ at both positive and negative
biases on the \textsl{poly}-\ce{Si} electrode.
These results are in consistent with the literature
\cite{SiO2:FN:Barriers:JAP53:5052:1982, Nasyrov:JAP105:123709}

\begin{figure}[tb]
	\includegraphics[width=\columnwidth]{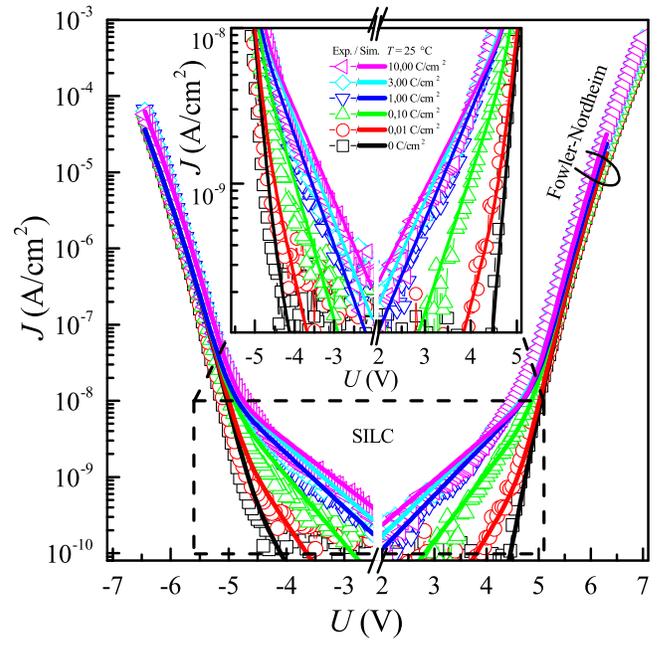}
	\caption{(Color online)
		Experimental (characters) and
		calculated (lines)
		current-voltage characteristics
		of $n$-\ce{Si}/\ce{SiO2}/\textsl{poly}-\ce{Si} structures
		at room temperature ($25$\,\celsius\xspace)
		after different SILC stress.
		Calculations were done using Eq.~(\ref{e:J:total}).
		The inset: SILC mode.}
	\label{f:IV25C}
\end{figure}

The experimental $J$-$V$ characteristics after SILC ($10$\,C/cm$^2$)
are shown in Fig.~\ref{f:IV25:annealed}
by round symbols.
At high bias $\abs{U}> 5$\,V on the \textsl{poly}-\ce{Si} contact,
the current through \ce{SiO2} films
is limited by Fowler-Nordheim tunneling.
The current at lower voltages $\abs{U}< 5$\,V is controlled by SILC.
The characters in Fig.~\ref{f:IV25C} represent the experimental
$J$-$V$ characteristics of
$n$-\ce{Si}/\ce{SiO2}/\textsl{poly}-\ce{Si} structures
at room temperature after different SILC stresses.
The SILC currents ($2.5$\,V$<\abs{U}<5$\,V) grow with
the increasing SILC stress.
The slope of $J$-$V$ curves at low voltages decreases
if the SILC is increased.
The temperature increase leads to
growing SILC currents on $30$--$50\%$,
but Fowler-Nordheim tunneling.

\begin{figure}
	\includegraphics[width=\columnwidth]{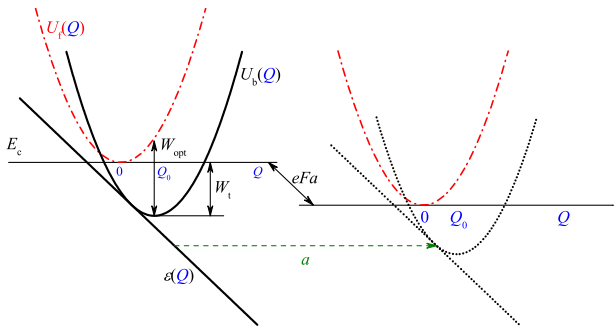}
	\caption{(Color online)
		Configuration diagram for two phonon-coupled traps.
		$U_\mathrm{f}(Q)$ is the potential energy of an empty oscillator (without trapped electron);
		$U_\mathrm{b}(Q)$ is the potential energy of an occupied oscillator (with trapped electron);
		$\varepsilon(Q)$ is the position of the energy level of the trapped electron dependent on coordinate $Q$;
		and $E_\mathrm{c}$ is the conduction-band edge.
		The solid and dotted lines refer to the initially occupied and empty state, respectively.
		The most probable tunneling transition for the electron
		when both oscillators take position $Q_0/2$
		is shown by the horizontal dashed arrow.
	}
	\label{f:PAT-diagram}
\end{figure}

When the trap density is high and the distance between them is short enough,
trapped electrons can tunnel between the neighboring traps
without ionization to the conduction band \cite{JAP:109:093705}.

A diagram of the electron tunneling from
a phonon-bound trap to the other one at a distance of $a$
in the presence of an external electric field
is shown in Fig.~\ref{f:PAT-diagram}.
The energy dependencies from generalized coordinate $Q$
of a system trapped-electron-bound-with-phonon
are shown by $U_\mathrm{b}(Q)$ terms.
The $U_\mathrm{f}(Q)$ terms correspond to ``free'' electrons
in the conduction band.
Solid lines represent the initial, before the tunneling state,
dashed lines represent the final state after tunneling.
Due to the external electric field
electrons on the neighboring traps have different energy levels
(slanted lines $\varepsilon(Q)$),
and the tunnel transition must be accompanied by inelastic processes,
like phonon emission and phonon absorption,
in order to compensate the energy difference.
The phonon-assisted tunneling model \cite{JAP:109:093705}
takes into account this circumstance.
According to this model, the rate of such transitions
is given by
\begin{equation}
	\begin{split}
		P_\mathrm{tun}=& \frac{2\sqrt{\pi}\hbar \Wt}{\m a^2 Q_0\sqrt{kT}}
		\exp(-\frac{2a\sqrt{2\m \Wt}}{\hbar})\times\\
		\times&\exp(-\frac{\Wopt-\Wt}{2kT})
		\sinh(\frac{eFa}{2kT}).
	\end{split}
	\label{e:P:PAT}
\end{equation}
Here 
$\hbar=h/2\pi$,
$Q_0=\sqrt{2(\Wopt-\Wt)}$,
$\Wt$ and $\Wopt$ are thermal and optical trap energies,
$k$ in the Boltzmann constant,
$T$ is the temperature.
At high temperatures, the charge transport through
the dielectrics is described within monopolar model involving
Shockley-Read-Hall equations and the Poisson equation:
\begin{equation}
	\pdv{\nt}{t}= \frac{1}{e}\divergence \vb*{J}
	  -a\divergence\left(\nt \left(1-\frac{\nt}{N} \right)P_\mathrm{tun} \frac{\vb*{F}}{\abs{\vb*{F}}}\right),
	\label{e:ShRH}
\end{equation}
\begin{equation*}
	\divergence \vb*{F} =-\frac{e\nt}{\varepsilon\varepsilon_0},
	\label{e:Poison}
\end{equation*}
where
$\nt$ is the filled trap density,
$t$ is time,
$N = a^{-3}$ is the trap density,
$J$ is the current density,
$\varepsilon$ is the static \ce{SiO2} permittivity ($\varepsilon=3.9$),
$\varepsilon_0$ is the electric constant.
In the static one-dimensional case,
Eq.~(\ref{e:ShRH}) gives the current-voltage characteristics
\begin{equation}
	J_\mathrm{tun} = \frac{e}{a^2}\frac{\nt}{N}\left(1-\frac{\nt}{N}\right)P_\mathrm{tun}.
	\label{e:J:tun}
\end{equation}
Note, the tunneling current takes a maximum value at $\nt/N=1/2$.
Taking into account Fowler-Nordheim tunneling (\ref{e:J:Fowler-Nordheim}),one 
can get the total current-voltage characteristics
\begin{equation}
	J = \frac{s}{S}J_\mathrm{tun} + J_\mathrm{FN},
	\label{e:J:total}
\end{equation}
where $J$ is the total current density,
$S$ is the total sample square ($8\times 10^4$\,$\mu$m$^2$),
$s$ is the square of SILC area.
We assume that charge stress induces extra traps
on a part of the \ce{SiO2} layer. This stressed part,
having effective square of $s$,
shunts the whole sample, and
gives additive to the current.

The comparison of experimental data (characters)
with the theory (\ref{e:J:total}) (lines)
is presented in Fig.~\ref{f:IV25C}.
As a result, we obtained thermal and optical trap energies
in \ce{SiO2} $\Wt = 1.6$\,eV and $\Wopt = 3.2$\,eV, respectively.
Simulations of experimental current-voltage characteristics
measures at $70$\,\celsius\xspace gives the same trap energy values.
Simulations within well-known Frenkel model
of isolated Coulomb center ionization \cite{Frenkel:Breakdown:1938:en,PhysRev:54:647}
predict that dinamic permittivity increases with the charge
stress up to $\varepsilon_\infty = 30$,
that is physically incorrect results.
All above confirm that using the model of
the phonon-assisted tunneling between traps
is correct.
The obtained model parameter values including
trap densities $N$ and
the squares of SILC area $s$ depending on
the total SILC charge $\Sigma$ are given in Table~\ref{t:params}.

We should note that the multi-phonon charge transport model
in the SILC mode was introduced earlier
\cite{SiO2:SILC:JAP86:2095:1999, SiO2:TAIT-SILC:JAP90:3396:2001}.
This model is not analytical, and it requires complex numerical
calculations to describe the SILC transport.

\begin{table*}
	\caption{Obtained values of parameter
		(\ref{e:J:Fowler-Nordheim}), (\ref{e:P:PAT}), (\ref{e:J:tun}), (\ref{e:J:total}).}
	\begin{tabular}{c|cccccc}\hline\hline
		$\Sigma$ (C/cm$^2$)&$0$&$0.01$&$0.1$&$1$&$3$&$10$\\ \hline
		\Wt (eV)&\multicolumn{6}{c}{$1.6$}\\
		\Wopt (eV)&\multicolumn{6}{c}{$3.2$}\\
		$\m/\me$&\multicolumn{6}{c}{$0.5\pm 0.02$}\\
		$N$ (cm$^{-3}$)&$\ll 10^{20}$&$1\times10^{21}$&$2\times10^{21}$&$4\times10^{21}$&$5\times10^{21}$&$7\times10^{21}$\\
		$s$ (nm$^{2}$)&$0$&$25\times 10^2$&$1\times 10^4$&$1.5\times 10^4$&$2\times 10^4$&$3\times 10^4$ \\
		\hline\hline
	\end{tabular}
	\label{t:params}
\end{table*}

We assume that oxygen vacancies (\ce{Si-Si} bonds)
act as traps responsible for SILC in \ce{SiO2},
which are generated as
$$\ce{#Si-O-Si# -> #Si-Si# + O},$$
where \ce{O} is the interstitial oxygen atom.
The electronic  structure of \ce{Si-Si} bonds in \ce{SiO2}
has been studied intensively
\cite{Defects:SiO2:PhysRevB.27.3780,SiO2:luminescence2.7eV:PhysRevLett.62.1388,a-SiO2:VO:luminescence:PhysRevLett.79.753,SiO2+SixNy:traps:CRSSMS36:129}.
The trap energies for electrons and holes,
calculated from the first principles (\ref{e:Wt:eh}),
are $1.2$\,eV and $1.6$\,eV, respectively.
The \ce{Si-Si} bond in silicon oxide acts as an amphoteric trap
capable of both electron and hole capturing.
Taking into account that such calculations usually
underestimate electron trap energy values,
one can conclude that the \textsl{ab initio}
simulation results agree to the thermal trap energy value
of $1.6$\,eV
obtained from transport simulations.

It is interesting to compare thermal trap energy $\Wt=1.6$\,eV
in \ce{SiO2} to the Stokes shift of \ce{Si-Si} defect luminescence.
It is known that \ce{Si-Si} bonds exhibit the ultraviolet
luminescence band with the photon energy of $4.4$\,eV
\cite{a-SiO2:VO:luminescence:PhysRevLett.79.753,a-SiO2:PL:PLR42:1765}.
The excitation maximum of this band is located at $7.6$\,eV
\cite{SiO2:luminescence2.7eV:PhysRevLett.62.1388,a-SiO2:VO:luminescence:PhysRevLett.79.753,a-SiO2:PL:PLR42:1765}.
A half of the Stokes luminescence shift $(7.6-4.4)/2=1.6$\,eV
is equal to the energy of electron and hole traps in \ce{SiO2},
which are \ce{Si-Si} bonds \cite{SiO2+SixNy:traps:CRSSMS36:129}.
This value is equal to the $\Wt$ value obtained from
charge transport  in \ce{SiO2} experiments in the SILC mode.
Thus, one can conclude that, namely, oxygen vacancies act as the
charge traps in \ce{SiO2} after the SILC stress.

Note that the obtained optical energy value \Wopt is twice bigger
than thermal trap energy \Wt. This empirical rule of the
multiphonon transport mechanism is executed for other dielectrics
\cite{Nasyrov:JAP105:123709,Novikov:APL94:222904, Al2O3:VO:jap108:013501,Perevalov:APL104:071904,HfO2:Transport:2014}.
To confirm or refute universal nature of this correlation,
it is necessary to collect statistics for a large number of dielectrics.
The optical trap energy $\Wopt=3.2$\,eV 
evaluated from the transport measurement
is very close to the electron trap energy of $3.1$\,eV
defined from the tunneling discharge of trapped in \ce{SiO2} electrons
\cite{Traps-SiO2:jap51:6258}.

Taking into account arguments above,
one can conclude that namely oxygen vacancies
are responsible for the SILC phenomenon,
but not other defects in \ce{SiO2}.

After the SILC stress some samples were treated by annealing
at the temperature of $250$\,\celsius\xspace during $120$ hours.
A comparison of the current-voltage characteristics measures for
the ``as deposited'',
after the SILC stress and after annealing structures
are shown in Fig.~\ref{f:IV25:annealed}.
One can see that $J$-$V$ curves, after annealing,
are almost identical to the ``as deposited'' control ones.
This phenomenon demonstrates that long-time annealing leads to
the electronic structure of the tunnel \ce{SiO2}, which is identical
to one of the ``as deposited'' films.
This means that the trap density decreases with annealing down to
initial values.
This phenomenon makes it possible to controllably suppress
the leakage currents caused by the tunnel \ce{SiO2} degradation
due to the SILC stress.

In conclusion, the electron transport due to the traps
in \ce{SiO2} was studied experimentally.
It was demonstrated that the transport is limited by
phonon-assisted tunneling between traps.
Comparison trap energy values obtained from different experiments on
charge transport in \ce{SiO2},
luminescence of oxygen vacancy (\ce{Si-Si} bond) in \ce{SiO2},
and \textsl{ab initio} simulations of \ce{Si-Si} bond electronic
structure,
it was found that the oxygen vacancies (\ce{Si-Si} bonds) act
as the charge localization centers (traps) in \ce{SiO2}.
The SILC stress leads to oxygen vacancies generation.
The thermal and optical trap energies were evaluated.
Long-time annealing at $250$\,\celsius\xspace
results in the recombination of oxygen vacancies and
interstitial oxygen,
and reduces the leakage current to the initial level.

This work was supported by the Russian Science Foundation
(grant \#16-19-00002).
The computations have been carried out
at the computational cluster in
the Rzhanov Institute of Semiconductor Physics SB RAS.

\bibliographystyle{apsrev4-1}
\newcommand{\bibpath}{../../../bibtex}
\bibliography{IEEEabrv,\bibpath/abrv,\bibpath/Technique,\bibpath/GeO2,\bibpath/TiO2,\bibpath/HfO2,\bibpath/SiO2,\bibpath/TaOx,\bibpath/Theory,\bibpath/percollation,\bibpath/Memristor,\bibpath/computing,\bibpath/ZrO2,\bibpath/Al2O3,\bibpath/minecite,\bibpath/FeRAM,\bibpath/high-k,\bibpath/Si3N4,\bibpath/Flash,\bibpath/Devices,silc}

\end{document}